\documentclass[11pt,a4paper]{article}

\usepackage{graphicx}
\usepackage{amsfonts}
\usepackage[margin=1in]{geometry}

\usepackage{latexsym,amssymb,amsmath,amsfonts,amscd,epsfig,amsthm,color,mathrsfs}
\usepackage{tikz}
\usepackage{hyperref}
\hypersetup{
    bookmarksnumbered=true, 
    unicode=false, 
    pdfstartview={FitH}, 
    pdftitle={Composing Quantum Algorithms}, 
    pdfauthor={Stacey Jeffery}, 
    pdfsubject={}, 
    pdfcreator={}, 
    pdfproducer={}, 
    pdfkeywords={}, 
    pdfnewwindow=true, 
    colorlinks=true, 
    linkcolor=blue, 
    citecolor=blue, 
    filecolor=blue, 
    urlcolor=blue 
}

\newcommand{\eps}{\varepsilon}

\def\e{\mathbf{e}}

\usepackage{hyperref}
\hypersetup{
    bookmarksnumbered=true, 
    unicode=false, 
    pdfstartview={FitH}, 
    pdftitle={Composing Quantum Algorithms}, 
    pdfauthor={Stacey Jeffery}, 
    pdfsubject={}, 
    pdfcreator={}, 
    pdfproducer={}, 
    pdfkeywords={}, 
    pdfnewwindow=true, 
    colorlinks=true, 
    linkcolor=blue, 
    citecolor=blue, 
    filecolor=blue, 
    urlcolor=blue 
}

\newcommand{\eq}[1]{\hyperref[eq:#1]{(\ref*{eq:#1})}}
\renewcommand{\sec}[1]{\hyperref[sec:#1]{Section~\ref*{sec:#1}}}
\newcommand{\thm}[1]{\hyperref[thm:#1]{Theorem~\ref*{thm:#1}}}
\newcommand{\lem}[1]{\hyperref[lem:#1]{Lemma~\ref*{lem:#1}}}
\newcommand{\cor}[1]{\hyperref[cor:#1]{Corollary~\ref*{cor:#1}}}
\newcommand{\app}[1]{\hyperref[app:#1]{Appendix~\ref*{app:#1}}}
\newcommand{\tab}[1]{\hyperref[tab:#1]{Table~\ref*{tab:#1}}}
\newcommand{\defin}[1]{\hyperref[def:#1]{Definition~\ref*{def:#1}}}
\newcommand{\fig}[1]{\hyperref[fig:#1]{Figure~\ref*{fig:#1}}}
\newcommand{\clm}[1]{\hyperref[claim:#1]{Claim~\ref*{claim:#1}}}
\newcommand{\conj}[1]{\hyperref[conj:#1]{Conjecture~\ref*{conj:#1}}}
\newcommand{\rem}[1]{\hyperref[rem:#1]{Remark~\ref*{rem:#1}}}

\newcommand{\thmthm}[2]{\hyperref[thm:#1]{Theorem~\ref*{thm:#1}} and~\hyperref[thm:#2]{\ref*{thm:#2}}}
\newcommand{\lemlem}[2]{\hyperref[lem:#1]{Lemma~\ref*{lem:#1}} and~\hyperref[lem:#2]{\ref*{lem:#2}}}

\title{Composing Quantum Algorithms}
\author{Stacey Jeffery\footnote{{\tt jeffery@cwi.nl}. This article appears as the December 2024 Complexity Theory Column in SIGACT News~\cite{jeffery2024sigact}.
}\\
QuSoft, CWI \& University of Amsterdam}
\date{}

\begin{document}


\maketitle

\begin{abstract}
Composition is something we take for granted in classical algorithms design, and in particular, we take it as a basic axiom that composing ``efficient'' algorithms should result in an ``efficient'' algorithm -- even using this intuition to justify our definition of ``efficient.'' Composing quantum algorithms is a much more subtle affair than composing classical algorithms. It has long been known that zero-error quantum algorithms \emph{do not} compose, but it turns out that, using the right algorithmic lens, bounded-error quantum algorithms do. In fact, in the bounded-error setting, quantum algorithms can even avoid the log factor needed in composing bounded-error randomized algorithms that comes from amplifying the success probability via majority voting. In this article, aimed at a general computer science audience, we try to give some intuition for these results: why composing quantum algorithms is tricky, particularly in the zero-error setting, but why it nonetheless works \emph{better} than classical composition in the bounded-error setting.
\end{abstract}

\section{Introduction}

Algorithms designers and programmers make ubiquitous use of subroutines. Not only does this allow code to be easily reused, perhaps even in subroutine libraries used across platforms, it adds structure to a program, or algorithm, that makes the whole thing easier to analyze. For algorithms designers, the most important reason to use subroutines is probably that by black-boxing the algorithms designed by your ingenious colleagues, you can use them as building blocks within your own new algorithms. 

Composition is a fundamental idea in complexity theory as well. One of the motivations for using polynomials as the class of growth functions considered ``efficient'' is that we feel intuitively that whenever we compose ``efficient'' algorithms, the resulting algorithm should also be ``efficient''~\cite{arora2012ComputationalComplexity}. Perhaps one reason we take for granted that this should be true is the ease with which we can compose and analyze classical algorithms. If an algorithm ${\cal A}$ makes $Q$ queries to a subroutine ${\cal B}$ with time complexity $T({\cal B})$, and $L$ additional operations, then its complexity is:
$$T({\cal A}) = Q\cdot T({\cal B})+L,$$
as any first-year computer science student could tell you. This allows us to say things about the complexity of \emph{composed} functions. Let $f:\{0,1\}^{n}\rightarrow \{0,1\}$ be the \emph{composed function} $f=g\circ h$, where $g:\{0,1\}^m\rightarrow\{0,1\}$ for some $m=m(n)\leq n$, and $h:\{0,1\}^{n/m}\rightarrow\{0,1\}$, defined on $x\in \{0,1\}^n$ by
$$f(x)=f(x^{(1)},\dots,x^{(m)}) = g(h(x^{(1)}),\dots,h(x^{(m)})),$$
where each $x^{(i)}$ is a $n/m$-bit string. Let $\mathbf{D}(f)$ be the \emph{deterministic query complexity} of $f$, also called the \emph{decision tree complexity}, which is the minimum number of queries to the bits of the input needed to decide $f$ (on the worst-case input). Then we can immediately see that
$$\mathbf{D}(g\circ h)\leq \mathbf{D}(g)\cdot \mathbf{D}(h),$$
by simply composing query-optimal deterministic algorithms for $g$ and $h$ (that is, every time the algorithm for $g$ makes a query to the input, replace it with a subroutine call to $h$ on the appropriate part of the input). 

A similarly easy composition applies to the \emph{expected} running time of \emph{Las Vegas} algorithms, which are randomized algorithms that always output the correct answer, but might run forever: In a Las Vegas algorithm ${\cal A}$, for any fixed input $x$ (which we generally leave implicit), the running time $T({\cal A})$ of the algorithm is a random variable, and while its expected value (over the random choices made by the algorithm) is generally finite, the maximum value it may obtain could be unbounded.  
If ${\cal A}$ is a Las Vegas algorithm that makes $Q$ queries to a Las Vegas subroutine ${\cal B}$, and $L$ additional operations, then\footnote{We make the reasonable assumption that the number of queries, $Q$, is independent of the time the queries take, which would be true, for example, if ${\cal A}$ were designed with each subroutine query treated as a black box whose cost is always unit.}
$$\mathbb{E}[T({\cal A})]=\mathbb{E}[Q]\cdot \mathbb{E}[T({\cal B})]+\mathbb{E}[L].$$
Let $\mathbf{R}_0(f)$ denote the \emph{zero-error randomized query complexity of $f$}, which is the minimum expected number of input queries needed by any Las Vegas algorithm for $f$ (on the worst-case input). Then we can easily see that
\begin{equation}\label{eq:R_0-comp}
\mathbf{R}_0(g\circ h)\leq \mathbf{R}_0(g)\cdot \mathbf{R}_0(h).
\end{equation}

This naive composition doesn't quite work if we use randomized algorithms that have some probability of error. Suppose $\widetilde{\cal B}$ is a \emph{bounded-error} (or Monte Carlo) algorithm, meaning that it is a randomized algorithm that outputs the correct answer with probability at least $2/3$ on every input, and, in contrast to Las Vegas algorithms, for every input $x$, there is some finite number of steps after which it is guaranteed to have halted -- we call this the running time, and it is not a random variable. Note that Monte Carlo algorithms are generally easier to come by than Las Vegas algorithms, since you can turn any Las Vegas algorithm with worst-case expected running time $T$ into a Monte Carlo algorithm with (deterministic) running time $O(T)$ by simply stopping the Las Vegas algorithm after $3T$ steps. By Markov's inequality, the probability the Las Vegas algorithm has not yet found the correct answer is at most $1/3$.

While $2/3$ is generally considered an acceptable correctness probability, in the context of a subroutine that is called $Q$ times by an algorithm ${\cal A}$ that expects the \emph{right} answer, this is insufficient. Assuming $Q\gg 1$, with overwhelming probability, one of the calls to $\widetilde{\cal B}$ will return the wrong answer, in which case, we no longer have any guarantees on the correctness of ${\cal A}$. To ensure that with probability at least $2/3$, all subroutine calls return the correct answer, we need the probability of error on each subroutine call to be $\ll \frac{1}{Q}$, so we can apply a union bound. 
The standard way to achieve this is as follows: every time ${\cal A}$ wants to call the subroutine, we will make about $\log Q$ parallel calls to $\widetilde{\cal B}$, and return the most commonly seen answer (the majority)\footnote{Note that majority voting only works if the desired behaviour of $\widetilde{\cal B}$ is to output a deterministic bit or string, as opposed to a distribution, or an element from a set of many possible correct answers.}. Then, by a binomial tail bound, each of the $Q$ queries made by $\cal A$ is correct except with probability $\ll\frac{1}{Q}$, as needed. If the resulting algorithm is called $\widetilde{\cal A}$, we have:
\begin{equation}\label{eq:bd-comp-rand}
T(\widetilde{\cal A}) = O\left(Q\cdot T(\widetilde{\cal B})\log{Q}+L\right).
\end{equation}
Letting $\mathbf{R}_2(f)$ denote the \emph{bounded-error randomized query complexity} of $f$, which is the minimum query complexity of a bounded-error randomized algorithm that decides $f$, we then have:
\begin{equation}
\mathbf{R}_2(g\circ h)\leq O\left(\mathbf{R}_2(g)\cdot \mathbf{R}_2(h)\log m\right).
\label{eq:R_2-comp}
\end{equation}
This inequality is tight: there exist composed functions for which this log factor is necessary~\cite{blais2019separation}.

In complexity-theoretic terms, this gives us, for deterministic, zero-error, and bounded-error algorithms respectively:
$${\sf P}^{\sf P} = {\sf P}, \qquad {\sf BPP}^{\sf BPP} = {\sf BPP}, \qquad\mbox{and}\qquad {\sf ZPP}^{\sf ZPP} = {\sf ZPP},$$
(all of which relativize), meaning that each of these three classes is \emph{self-low}. These complexity-theoretic statements are, of course, quite granular compared to the finer-grained complexities of composition described above, but they roughly state that each of the three types of algorithms -- efficient deterministic, efficient Monte Carlo, and efficient Las Vegas -- are ``closed under composition'', as one would intuitively hope.

All of this is quite elementary, and I don't mean to be a bore. My point in saying all of this is to set up a contrast with composition in \emph{quantum} algorithms, which is \emph{not} so simple. 
The first indication of this, to my knowledge, was a result by Buhrman and de Wolf in 2002 showing that a quantum analogue of \eq{R_0-comp} does \emph{not} hold~\cite{buhrman2003zeroerror}. They showed an oracle relative to which
$${\sf ZQP}^{{\sf ZQP}}\neq {\sf ZQP}.$$
One reason that this is surprising is that \emph{bounded-error} quantum query complexity is known to compose nicely. 
There is nothing that prevents a quantum algorithm from calling a subroutine, and an analogue of \eq{R_2-comp} for quantum algorithms is quite straightforward. But more than that, it is possible to show an even stronger statement, \emph{without} log factors:
\begin{equation}\label{eq:no-log-queries}
\mathbf{Q}_2(g\circ h)\leq O(\mathbf{Q}_2(g)\cdot\mathbf{Q}_2(h)),
\end{equation}
where $\mathbf{Q}_2(f)$ is the minimum quantum query complexity of any bounded-error quantum algorithm for $f$~\cite{reichardt2009span}.

We will return to this way in which quantum composition works \emph{better} than classical shortly. For now, we focus on the \emph{difficulty} of composing quantum algorithms, which stems from (at least) two issues:
\begin{enumerate}
\item Whereas classical algorithms may call a distribution of different subroutines, the quantum analogue is a \emph{superposition} of different subroutines. Unlike a classical distribution, which can be visualized as a tree, quantum branches of a superpostion cannot be considered as separate non-ineracting branches forever. We generally visualize them as coming back together for the next operation -- indeed, unlike randomness, it is possible to evolve a superposition supported on many states back to a point function supported on a single state. So, naively, one must wait for all branches of the superposition to terminate before the next operation can be applied.
\item The quantum analogue of probability is \emph{amplitude}, and it can be negative. Nonzero amplitude on an accepting state may later be cancelled by negative amplitude, in contrast to nonzero probability on an accepting state, which, in a zero-error classical algorithm, is definitive evidence that ``accept'' is the correct outcome.
\end{enumerate}
The second issue is severe, and is why zero-error quantum algorithms do not compose, which we discuss more in \sec{zero}. The first issue turns out to be surmountable, if we simply look at quantum algorithms the right way. 

To illustrate the first issue, imagine having (classical) subroutines ${\cal B}_1,\dots,{\cal B}_N$ called by an algorithm $\cal A$, 
 and let $Q_i$ be the number of calls to ${\cal B}_i$. Then clearly
\begin{equation}\label{eq:rand-exp}
\mathbb{E}[T({\cal A})]=\sum_{i=1}^N\mathbb{E}[Q_i]\mathbb{E}[T({\cal B}_i)]+\mathbb{E}[L],
\end{equation}
where, as usual, $L$ is the number of additional operations used by $\cal A$. For simplicity, suppose the total number of subroutine queries made by $\cal A$, $Q$, is deterministic, and let $p_{j,i}$ be the probability that the $j$-th subroutine call is to ${\cal B}_i$. Then we can rewrite this:
\begin{equation}\label{eq:rand-avg}
\mathbb{E}[T({\cal A})]=\sum_{i=1}^N\sum_{j=1}^Qp_{j,i}\mathbb{E}[T({\cal B}_i)]+\mathbb{E}[L].
\end{equation}
An analogous statement for quantum algorithms is not obvious, and even in the case where the running times $T({\cal B}_i)$ are deterministic (which does simplify things for quantum algorithms), 
naive composition yields a statement analogous to:
\begin{equation}
\mathbb{E}[T({\cal A})] = Q\max_{i\in [N]}T({\cal B}_i)+\mathbb{E}[L].
\end{equation}
This is because, for the most part, quantum algorithms have been viewed in the \emph{quantum circuit model}, which treats them in analogy to classical deterministic algorithms, where each operation is something to be done once, at a fixed time. If the $j$-th subroutine call is something that happens once, at a fixed time, then it certainly can't happen before the $(j-1)$-th subroutine call has terminated (in all branches of superposition), and so the algorithm must wait $\max_{i\in [N]} T({\cal B}_i)$ time steps, assuming each subroutine has a non-zero probability of being called in the $(j-1)$-th query.

In contrast, randomized algorithms are often viewed as graph-like structures, with probability flowing through them, and so a particular operation might be applied at some step with some probability, and at a later step with some other probability. Taking a similar view of quantum algorithms turns out to make composition much more straightforward. While we can't hope to prove a quantum statement analogous to \eq{rand-exp} for ${\cal A}$ a Las Vegas algorithm\footnote{In fact, \emph{quantum Las Vegas complexity}, is better defined in a more natively quantum way, as in \cite{belovs2023LasVegas}, on which \cite{belovs2024transducers} is partially based.}, due to the result of Buhrman and de Wolf~\cite{buhrman2003zeroerror}, we do have the next best thing: a quantum analogue of \eq{rand-avg} for $\cal A$ a  \emph{bounded-error} quantum algorithm~\cite{jeffery2022subroutines}.
This quantum analogue does not use naive quantum composition to obtain $\cal A$, but a more involved way of composing quantum algorithms, viewing them as quantum analogues of random walks. The results of \cite{jeffery2022subroutines} apply only to subroutines that compute a single bit, but in \cite{belovs2024transducers} we generalize this to quantum subroutines that compute any \emph{unitary map} (the kind of map quantum computations are capable of), using a new quantum algorithmic model that generalizes quantum random walks, called \emph{transducers}. 
In \sec{q-vs-r}, we discuss in more detail why quantum composition is different from classical composition, but nonetheless works, if we are satisfied with getting a bounded-error algorithm. This is done using pictures, rather than equations, so it is easy to follow on an intuitive level, but does not contain enough detail for a rigorous understanding. For that, the reader may refer to \cite{jeffery2022subroutines} or \cite{belovs2024transducers}. 

I'd like to return your attention to \eq{no-log-queries}, where we seem to magically save a log factor. Recall that the log factor repetition
in composed classical bounded-error algorithms comes from repeating each subroutine call $\log{Q}$ times, in order to decrease the error of each subroutine query to around $\frac{1}{Q}$ via majority voting. Are quantum algorithms somehow able to avoid this repetition? 

For a while I was doubtful that avoiding the log-fold repetition was possible. The result in \eq{no-log-queries} is proven non-constructively, so it could be that any algorithm for $f=g\circ h$ achieving the optimal query complexity $O(\mathbf{Q}_2(g)\mathbf{Q}_2(h))$ does have some kind of log repetition, it's simply recycling queries, somewhat like randomness can be recycled in majority voting to avoid a log-factor overhead in the random bits needed in error-reduced algorithms~\cite{impagliazzo1989recyleRandomBits}. 

However, it turns out quantum algorithms really can avoid repeating subroutines log-many times: we were able to show that if ${\cal A}$ is a bounded-error quantum algorithm that makes calls to a subroutine, whose desired functionality is implemented with bounded error by another quantum algorithm $\widetilde{\cal B}$, then there is a quantum algorithm $\widetilde{\cal A}$ that implements the desired behaviour of ${\cal A}$ in complexity~\cite{belovs2024transducers}:
$$T(\widetilde{\cal A}) = O\left(Q\cdot T(\widetilde{\cal B})+L\right),$$
where, as usual, $L$ is the number of additional operations made by $\cal A$, giving a quantum analogue of \eq{bd-comp-rand}, but without the log factor. A similar log-factor-free statement applies when there are many different subroutines implemented with bounded error, in which case we get a complexity analogous to \eq{rand-avg}, up to constant factors. 

In order to prove this, we make use of our new formalism for quantum algorithms, the aforementioned \emph{transducers}. If an algorithm has some error with respect to its desired behaviour, it will map to a \emph{perturbed} (with respect to the desired behaviour) transducer. Just as an algorithm's error can be reduced to any $\eps$ at a $O(\log \frac{1}{\eps})$ overhead using majority voting, a transducer can be \emph{purified} to a transducer with arbitrarily small perturbation $\delta$, but unlike majority vote, purification results in a $O(1)$ multiplicative overhead, \emph{independent of $\delta$}. Whenever we compile transducers back into algorithms, a constant error $\eps$ is introduced, resulting in a bounded-error algorithm, so this does not give us a way to turn a bounded-error quantum algorithm into a $o(1)$-error quantum algorithm with just $O(1)$ overhead. However, it does give a way to compose without log factors, by purifying all subroutines to have arbitrarily small perturbation, so that when the transducers are composed to get a new transducer, the cumulative perturbation is still $o(1)$, and then turning back into a bounded-error quantum algorithm. 

I'm not really a complexity theorist -- though I am friends with several -- so I am not entirely sure where they stand on log factors, but it seems they are not too bothered by them -- after all, they're not even polynomial! To be honest, as an algorithms person, me neither, for the most part.\footnote{They can pose a rather more severe problem when composing algorithms to non-constant depth.} In analyzing algorithms, I'm wont to hide log factors -- even polylog factors -- in my asymptotics, using the notation $\widetilde{O}(\cdot)$. However, I think it is still quite striking that these log factors need not appear in composed quantum algorithms. 
I think it may give us some insight into the power of quantum computers, although I don't yet know what that insight is, and won't be telling you about it here. But in \sec{log}, I will discuss this result more, and give some idea of where it comes from using a toy example.

\paragraph{Note on the model of computation:} When we talk about classical computation, we assume the word RAM model, in which random access memory reads and writes (to our working memory, and perhaps a memory storing a description of the program to be executed) have unit cost, as do basic operations, such as addition, of words of a size large enough to address this memory. Our quantum model will be the quantum analogue of this. The quantum word RAM model is less standard than its classical counterpart, partially because implementing a quantum random access gate might be much more difficult than quantum gates acting only on one or two qubits (quantum bits). We will not concern ourselves with the question of how realistic this model of quantum computation is in the near-term, as our main interest here is theoretical comparison between classical and quantum computation, so we give the same powers to both models. Ref.~\cite{belovs2024transducers} gives some weaker composition results that also apply in the strict quantum circuit model, in which random access gates are not consider a basic operation. 

\section{Quantum vs. Randomized Algorithms}\label{sec:q-vs-r}

The goal of this section is to illustrate the difficulty of composing quantum algorithms, and specifically, to convey the following:
\begin{enumerate}
\item Composing quantum algorithms to achieve a complexity that is a quantum analogue of \eq{rand-avg} is not obvious.
\item It is nonetheless possible, with the caveat that the composed algorithm will have bounded error.
\end{enumerate}
We also hope to lay the foundations for seeing why zero-error composition is \emph{not} possible, which we discuss more in \sec{zero}.
For simplicity, throughout this section, the reader may assume that the quantum subroutines have no error, and ${\cal B}_i$ terminates after a fixed time $T({\cal B}_i)$, though this time may vary in $i$. 
Towards these goals, in this section we describe visualizations of randomized and quantum algorithms that allow us to compare them at an intuitive level without  the use of math\footnote{Math is still very much recommended for actually proving things. For details, see \cite{jeffery2022subroutines,belovs2024transducers}.}. 

\begin{figure}
\centering
\includegraphics[scale=.5]{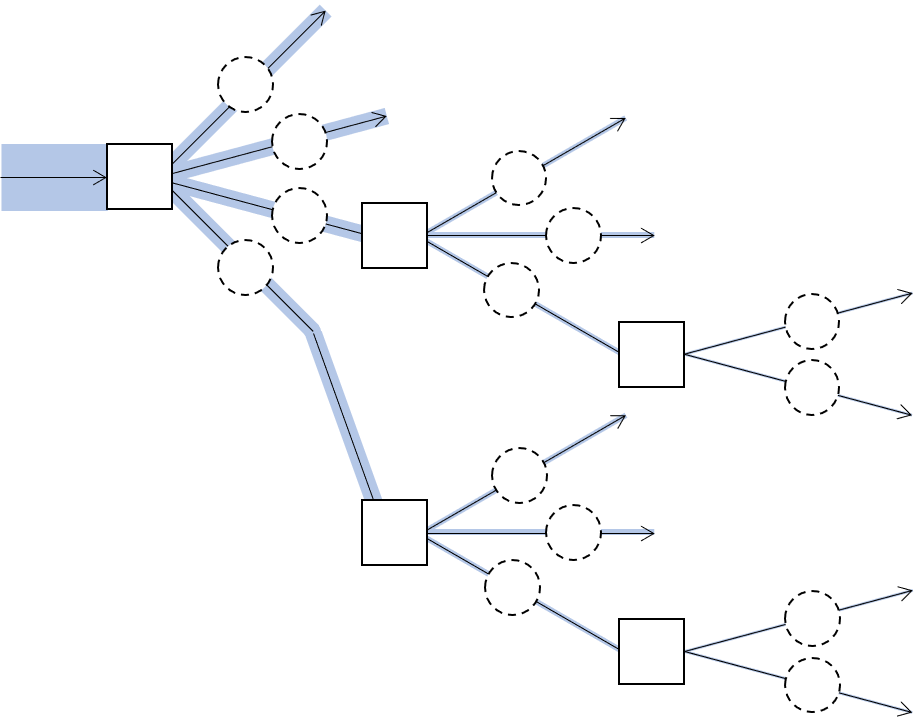}
\caption{A run of a randomized algorithms, visualized as probability flowing through a tree. The incoming arrow is the entrance, and outgoing arrows are terminals.}\label{fig:tree}
\end{figure}

\paragraph{Randomized algorithms, as trees} A randomized computation can be visualized as a tree (see \fig{tree}). 
There is an \emph{entrance} at the root, and outgoing \emph{terminals} at each leaf, each of which is either \emph{accepting} or \emph{rejecting}.
The internal tree nodes, shown as rectangles, represent coin flips, and some extra nodes on the edges might represent deterministic computations. We use dashed lines to represent a computation that we model as an oracle call, either querying the input directly, or perhaps meant to be instantiated by a subroutine. 

Of course, we sometimes visualize randomized computations as another type of graph, like a DAG, or more general random walk. However, as long as we don't care about the amount of memory the algorithm uses, we can always model it as a tree, indicating that we remember every coin flip ever made (even if our future computations might not depend fully on them). So the tree picture is without loss of generality, as far as time complexity is concerned.

Importantly, in our visualizations, the node you are in will not encode the full state of the algorithm, but simply where you are in the computation. We assume a ``walker'' moving through the tree carries along some extra information as well, based on queries made and computations performed so far.
As such a tree's structure may be adaptive, depending on the input via results of computations in the dashed circles. 
To visualize a run of the computation on a particular input, we can imagine that we have a unit of probability on the entrance, and as we run the algorithm, it flows through the tree, from left to right, splitting when it gets to an internal tree node (evenly, assuming coin flips are unbiased), until it gets to a terminal. It can be helpful to think of this probability as water flowing through the graph, whose edges are like pipes, possibly carrying with it some information about the algorithm's current state.\footnote{If you want to take this metaphor to an obscene level, you can think of the flow as consisting of different types of liquids that don't mix, like water and oil, whose volumes add up to the total flow. The type of liquid encodes the additional information.} 
The amount of water already at the terminals after $\ell$ steps -- equivalently, the amount of water at terminals with distance at most $\ell$ from the entrance -- is precisely the probability that the algorithm has halted after at most $\ell$ steps. \fig{tree} also shows this \emph{probability flow}, which includes the probability on the edges at \emph{all} steps of the algorithm, rather than only at a particular step.\footnote{So if you add up all probabilities shown in the probability flow, you will not get 1, but rather, you will get the expected running time of the algorithm.}

It is quite intuitive from \fig{tree} that if we instantiate the dashed circles with programs of varying length -- for simplicity, imagine deterministic programs, which are just line graphs -- we get a complexity (expected distance from entrance to root) as in \eq{rand-avg}.

\paragraph{Quantum algorithms are not trees} We can view quantum computations in a somewhat similar picture, but we replace the concept of randomness with its quantum analogue, \emph{superposition}, and probability with its quantum analogue \emph{amplitude}. A quantum state is a vector, not unlike a probability distribution, but a quantum state is $\ell_2$-normalized instead of $\ell_1$-normalized like a probability distribution. The squared norms of the amplitudes, which sum to 1, are actually probabilities. 

Unlike probability, amplitude can be negative (or even complex), and -- actually because of this -- in contrast to a random variable, a superposition has no entropy. It's possible to move from a ``classical deterministic'' state, such as $0^n$, into a superposition of many states, and then back to a ``classical deterministic'' state. So we cannot generally visual a quantum algorithm as a tree. Instead, it looks more like the diagram below\footnote{For readers used to quantum circuit diagrams: the lines do \emph{not} represent different qubits, but rather, different states, so in an $n$-qubit system, there are $2^n$ of them.}:

\begin{center}
\includegraphics[scale=.5]{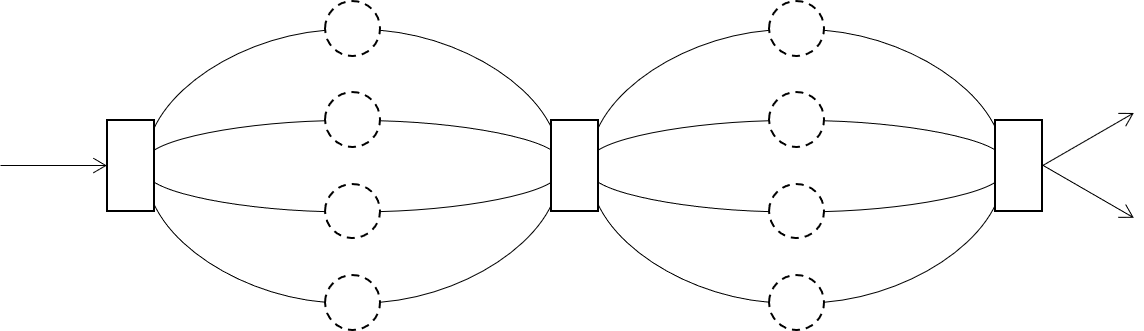}
\end{center}

\noindent Each of the rectangles represents some quantum operation, and as above, the dashed circle represents a query to the input (or perhaps a subroutine). There is an accepting terminal, and a rejecting terminal. In this visualization, any $Q$-query quantum algorithm looks the same, so this basic picture tells us very little about the program's structure. To visualize a run of the algorithm, we can imagine its amplitude flowing through the graph, as we did with probabilities.

\begin{center}
\includegraphics[scale=.5]{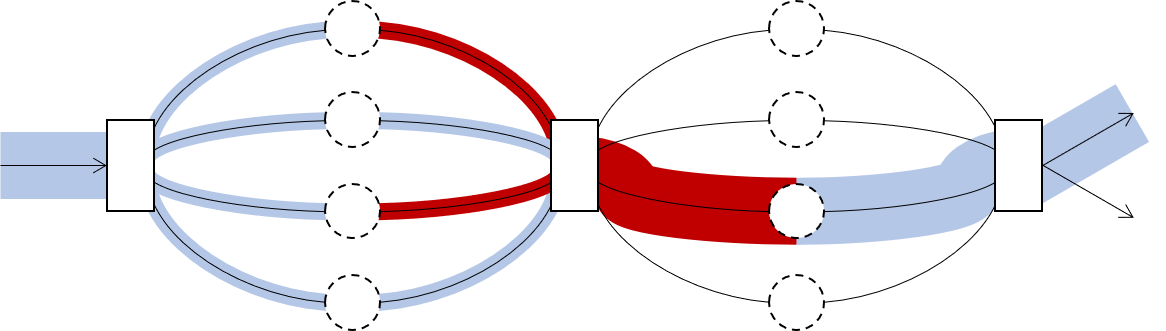}
\end{center}

When we show \emph{amplitude flows}, the strength of the flow will be visualized as the \emph{square} of the amplitude, so they all add up to 1 (on any entrance-terminals cut), whereas the colour will tell us the sign: blue for positive, and red for negative\footnote{In general, the amplitude could take any direction $e^{-i\theta}$ on the unit circle, but we will restrict to postive, $(+1)$ and negative, $(-1)$.}. Visualizing a quantum algorithm this way is, admittedly, much less helpful and intuitive than visualizing a randomized algorithm as a tree, but it does allow us to think about quantum composition. Imagine you want to instantiate the dashed circles with calls to different subroutines. A deterministic subroutine is just a line, so we get something like the following:

\begin{center}
\includegraphics[scale=.5]{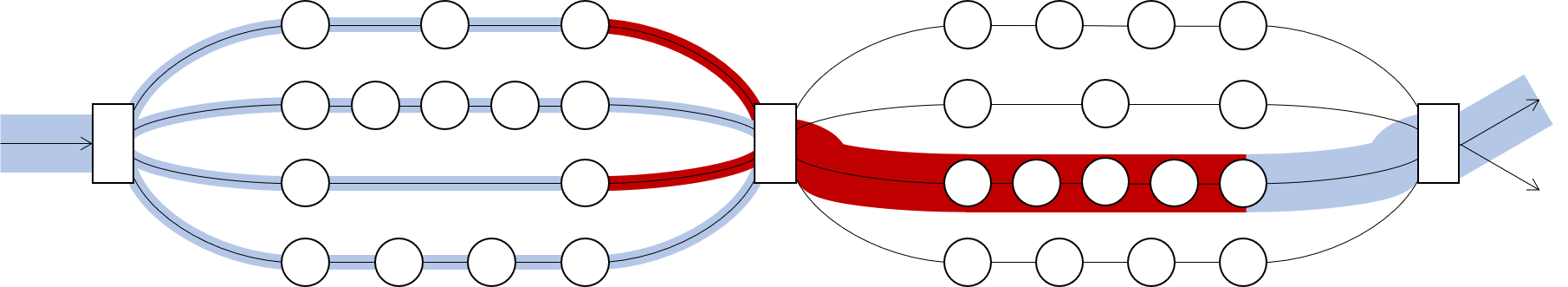}
\end{center}

\noindent Each circle above is a subroutine step. The number of subroutine steps varies in different branches of the superposition, just as the number of subroutine steps might vary in different tree branches in a randomized algorithm. 
What is the complexity of this composed quantum algorithm?  
To run this algorithm, first we apply the first quantum operation, represented by the first rectangle. Before we apply the second, it seems like we need to complete all subroutine calls -- consisting of the operations represented by circles -- because the second rectangle depends on all the inputs, so naively, we have to wait the maximum path length. This is certainly true if we think of the second operation as something we're going to apply once, at some fixed time, as we do in the most common model of quantum algorithms, the \emph{quantum circuit model}. 
This gives complexity:
\begin{equation}
T({\cal A})=Q\max_{i\in [N]} T({\cal B}_i)+L,\label{eq:quant-max}
\end{equation}
which, compared to \eq{rand-avg}, is terrible, as we've replaced an average with a maximum. This illustrates that it is not straightforward to have quantum subroutines that take different amounts of time in different branches of the superposition. It seems like we need to wait for the slowest branch of the superposition to terminate before we can move on to the next step, and until a couple of years ago, no better expression than \eq{quant-max} was known.

This issue doesn't come up in a tree, obviously, since paths never come together again. However, as previously mentioned, we often do visualize randomized algorithms as something other than a tree. What is the expected time to get from the entrance to one of the terminals in the following DAG (all edge directions are left-to-right), representing a randomized computation where we forget whatever randomness has been used so far?

\begin{center}
\includegraphics[scale=.5]{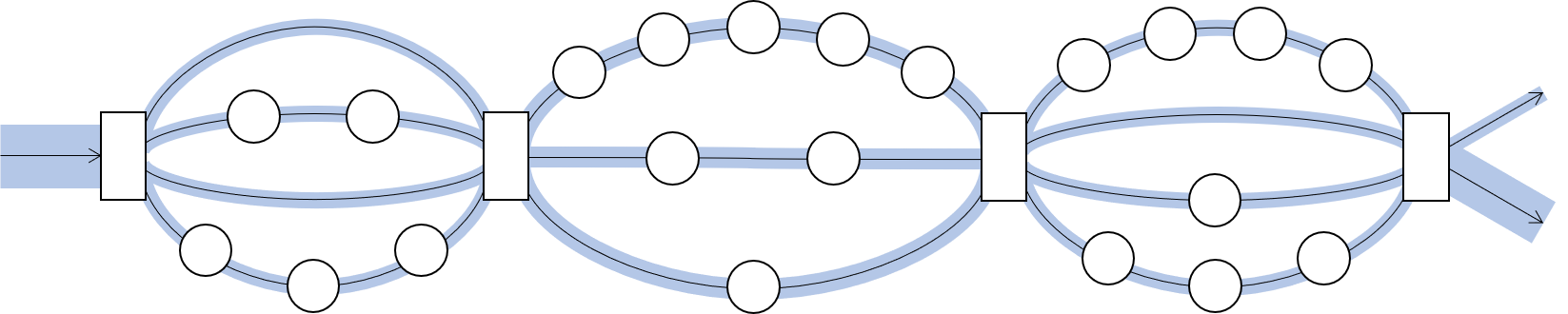}
\end{center}
It is obviously the expected path length. As ``liquid'' moves through the graph, it eventually all ends up at the terminals. It doesn't matter if you let liquid through the rectangles as it arrives, or wait for it all to arrive before distributing it out the other side, it will all end up at the terminals, carrying the same information either way, because probability can do nothing but add up.

It turns out that this intuition also holds (with somewhat more difficulty in making it precise enough to turn into a proof) in quantum algorithms. 
If we think of the quantum algorithm in analogy to a random process -- actually, as the quantum analogue of a random walk -- we find that it's fine if we send some amplitude through a box before the rest of the amplitude has caught up: it will all add up to the same thing on the terminals in the end, no matter what speed different parts of it get there.
Thus, if we combine the outer quantum algorithm, consisting of just the boxes and the dashed circles representing queries, with the quantum subroutines instantiating those queries -- 
in a particular way inspired by quantum random walks -- everything will add up correctly to the right state on the terminals after the final rectangle. And if we just wait until \emph{most} of the amplitude has made it through the final rectangle, we will end up with a state that is \emph{mostly} correct, giving a bounded-error quantum algorithm $\widetilde{\cal A}$ with complexity:
\begin{equation}
T(\widetilde{\cal A}) = O\left(\sum_{j=1}^Q\sum_{i=1}^Nq_{j,i}T({\cal B}_i)+L\right),\label{eq:quant-avg}
\end{equation}
where, $Q$ is the number of subroutine calls, and $q_{j,i}$ is the squared norm of the amplitude on the $i$-th superposition branch right before the $j$-th subroutine query. 
This was shown (with polylog factor overhead) for subroutines that compute a single bit in \cite{jeffery2022subroutines}, and later generalized to arbitrary (unitary) quantum subroutines in~\cite{belovs2024transducers}.
We can even replace the complexities $T({\cal B}_i)$ with expected complexities (we have not talked about how quantum algorithms can have a running time that is a random variable, but it is possible) to get a quantum analogue of \eq{rand-avg}.

We have been assuming, for simplicity that the subroutines have no error. If we instead have bounded-error subroutines, then we can use majority voting (assuming the desired behaviour of the subroutines is to output a deterministic string) to obtain the complexity in \eq{quant-avg}, but with logarthmic overhead. However, using \emph{purifiers}, we can even get this complexity expression with no log factors even when the subroutines have bounded error. We discuss this more in \sec{log}.

The reason quantum composition does not work in the zero-error case, that is, if we want the composed algorithm $\cal A$ to have zero error, is that we would need to wait for all amplitude to get to the end of the graph if we want to have no probability of error. In the next section, we will explore this subtle difference between quantum and classical composition.

\section{Impossibility of Zero-error quantum Composition}\label{sec:zero}

In~\cite{buhrman2003zeroerror}, Buhrman and de Wolf show that there is an oracle ${\sf A}$ such that
$$ {\sf ZQP}^{{\sf ZQP}^{\sf A}} \neq {\sf ZQP}^{\sf A},$$
by exhibiting a composed function $f=g\circ h$ for which $\mathbf{Q}_0(g)=1$, $\mathbf{Q}_0(h)=O(1)$, and $\mathbf{Q}_0(f) \geq \frac{m}{2}+1$. 
We describe this composed function, and try to give some intuition why its zero-error quantum query complexity is not also constant. 

\begin{figure}
\centering
\includegraphics[scale=.5]{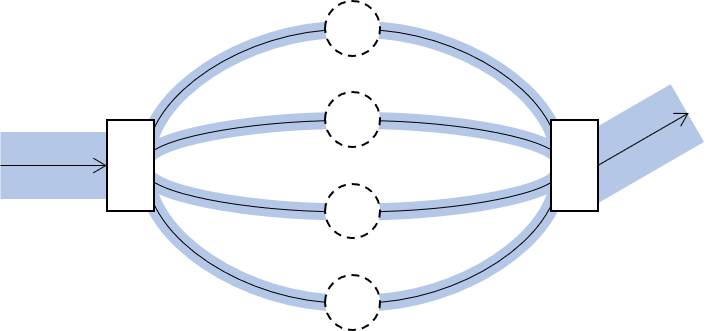}
\hspace{20pt}
\includegraphics[scale=.5]{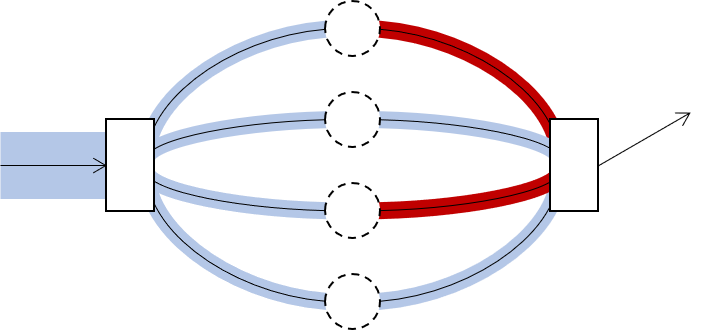}
\caption{On the left, we see a run of the algorithm on a constant input, and on the right, a run of the algorithm on a balanced input. In both cases, we first do a quantum Fourier transform (first box) to split one unit of amplitude uniformly into $m$ branches. In branch $i\in [m]$, we query $x_i$ (the dashed circle), changing the sign from positive (blue) to negative (red) if $x_i=1$. Finally, we invert the Fourier transform (last box), sending the sum of the amplitudes on all branches onto the accepting terminal. When $x$ is constant (left-hand side), all paths stay blue (positive), so they add up to a unit of amplitude going out the accepting terminal. When $x$ is balanced (right-hand side), half the paths become red (negative) and cancel with the blue paths, so they add up to 0 amplitude on the accepting terminal. Note that there are other implicit terminals (not shown), all of which are rejecting.}\label{fig:DJ}
\end{figure}

Let $|x|$ denote the Hamming weight of a binary string $x$. Let $g$ be the promise problem on the set $\{x\in \{0,1\}^m:|x|\in \{0,m/2\}\}$ for even $m$, defined:
$$g(x)=\left\{\begin{array}{ll}
1 & \mbox{if }|x|=0\\
0 & \mbox{if }|x|=m/2.
\end{array}\right.$$
This function decides if a string is \emph{constant} $0^m$; or \emph{balanced}, meaning it has the same number of 0s and 1s. It is well known in the quantum algorithms literature, because there is a quantum algorithm due to Deutsch and Jozsa~\cite{deutsch1992algorithm} -- one of the very first quantum algorithms in fact -- that outputs the correct answer using a single query to the input. So in particular, $\mathbf{Q}_0(g)=1$ (in fact, we have the stronger statement $\mathbf{Q}_E(g)=1$, where $\mathbf{Q}_E$ is the \emph{exact quantum query complexity}, since 1 is not merely the expected complexity, but an upper bound on the complexity). In contrast, if the input is constant, then a classical algorithm needs to see $m/2+1$ bits to be certain that they are all the same, so ${\sf R}_0(g)=m/2+1$. 

The Deutsch-Jozsa algorithm, which decides $g$ exactly with a single quantum query, relies precisely and crucially on the promised structure of the problem, which we can illustrate in our intuitive graph picture, in \fig{DJ}. The algorithm first does something called a \emph{quantum Fourier transform}, which maps $0^{\log m}$ to a uniform superposition over all $\log m$-bit strings -- a bit like flipping $\log m$ (quantum) coins -- representing each $i\in [m]$. Next, in each branch of the superposition, the algorithm queries the $i$-th bit. It does a pretty quantum thing with that bit: if the bit is 1, it changes the sign of the amplitude from positive to negative. That's shown as the colour changing from blue to red. Finally, it does something called an inverse Fourier transform, and the only thing we need to know about that is that the amplitude on $0^{\log m}$  coming out of that operation -- which is the accepting state/terminal (the rejecting terminal is not pictured, but all other states are rejecting) -- is the sum of all incoming amplitudes\footnote{This is actually not completely correct, but it gives the right mental picture if you're not following too closely. The actual amplitude on the accepting terminal is $1/\sqrt{m}$ times the sum of the $m$ incoming amplitudes, each of which is either $1/\sqrt{m}$ (blue) or $-1\sqrt{m}$ (red). The square roots are because quantum states are $\ell_2$-normalized.}. So one of two things happens: 
\begin{description}
\item[Constant (accepting) Case:] all the bits are 0, meaning all paths are blue, so they add up to unit amplitude (corresponding to probability 1) coming out the accepting terminal; or
\item[Balanced (rejecting) Case:] half the bits are 0 and half the bits are 1, meaning there are as many blue paths as red paths, and so they cancel, resulting in 0 amplitude (corresponding to probability 0) coming out the accepting terminal.
\end{description}
Note that if the Hamming weight were not exactly $m/2$ in the rejecting case, we would not get the perfect cancellation that makes this an exact, and thus a zero-error, algorithm.

\begin{figure}
\centering
\includegraphics[scale=.5]{h-tree}
\hspace{30pt}
\includegraphics[scale=.5]{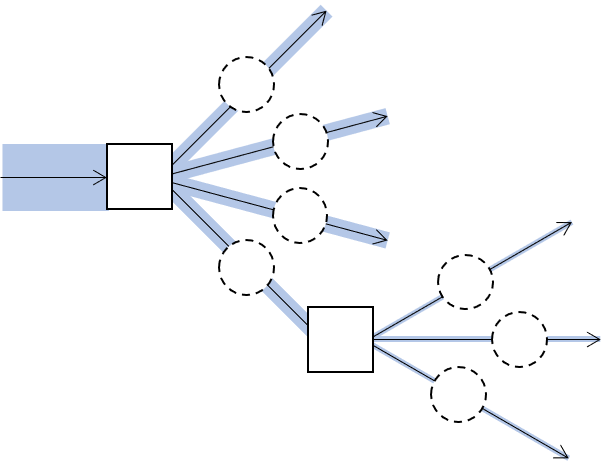}
\caption{Here we see the probability flow for two runs of a randomized algorithm for $h$ on different inputs: one with $|x|=m/2$ (left-hand side); and one with $|x|=3m/4$ (right-hand side). The dashed circles represent calls to a small subroutine that queries $x_j$ and $x_{m+j}$ in the $j$-th branch, and terminates if one of them is 1, in either an accepting terminal (if $x_{m+j}=1$) or a rejecting terminal (if $x_j=1$). In the left-hand image, which is the worst case, half the probability is already on terminals after one such step, and for the remaining half, a new $j'\in [m]\setminus\{j\}$ is chosen. The longest paths from the entrance to a terminal in the worst case are $m/2+1$ steps, but the total probability weight on these is very small.}\label{fig:h-tree}
\end{figure}

The inner function $h$ defining $f=g\circ h$ will also be a promise problem, and we will define it by its pre-images:
\begin{align*}
h^{-1}(1) & = \{0^mx^R: x^R\in\{0,1\}^m, |x^R|\geq m/2\}\\
h^{-1}(0) & = \{x^L0^m: x^L\in\{0,1\}^m, |x^L|\geq m/2\}.
\end{align*}
So we are promised that one half of the string is all-0s, and the other half has relatively many 1s, and we must decide which half is which. Even a classical algorithm can solve this with zero error in expected constant time: Alternatively query a random as-of-yet-unqueried bit $x_j$ on the left-half and, the corresponding $x_{j+m}$ on the right half until you see a 1, at which point, you're done. In the tree picture, this looks like \fig{h-tree}.

If you're maximally unlucky, you will query all the 0s in the half with 1s before you query a 1, meaning you will spend up to $2(m/2+1) = m+2$ queries, but this only happens with probability $\Theta(2^{-m})$. 
Since every two queries finds a 1 with probability at least $1/2$, the expected number of queries before a 1 is found is $O(1)$. 
Thus $\mathbf{Q}_0(h)\leq \mathbf{R}_0(h)=O(1)$.

We have seen that $\mathbf{Q}_0(g)$ and $\mathbf{Q}_0(h)$ are both $O(1)$, so, in light of the fact that $\mathbf{R}_0(g\circ h)\leq\mathbf{R}_0(g)\cdot \mathbf{R}_0(h)$, we might expect $\mathbf{Q}_0(g\circ h)$ to be $O(1)$, but in fact, we have~\cite{buhrman2003zeroerror}:
$$\mathbf{Q}_0(g\circ h)\geq m/2+1.$$

The classical result $\mathbf{R}_0(g\circ h)\leq\mathbf{R}_0(g)\cdot \mathbf{R}_0(h)$ is based on the observation that we can simply run an optimal Las Vegas algorithm for $g$, and every time the algorithm tries to query the input, replace the query with a call to the optimal Las Vegas algorithm for $h$ (on the appropriate block of the input). Let us try to understand what goes wrong with quantum composition, by considering running the single-query Deutsch-Jozsa algorithm for $g$, but replacing the query with a call to our algorithm for $h$ (which we can turn into a quantum algorithm without much difficulty). 

To see the issue, consider an input $x=(x^{(1)},x^{(2)},x^{(3)},x^{(4)})$ to $g \circ h$ (so $m=4$) that induces a balanced input to $g$, 
meaning that exactly half of the strings $x^{(1)}$, $x^{(2)}$, $x^{(3)}$, and $x^{(4)}$ are 1-inputs to $h$. Each part, $x^{(i)}$, of $x$ is a $2m$-bit string with two parts, $x^{(i,L)}$ and $x^{(i,R)}$, one of which has weight 0, the other of which has weight at least $m/2$. Suppose further that $x$ has the following structure:
\begin{center}
\begin{tabular}{r|cc|cc|cc|cc}
 $x=$& $0^m$ & $x^{(1,R)}$ & $x^{(2,L)}$& $0^m$& $0^m$& $x^{(3,R)}$& $x^{(4,L)}$& $0^m$\\
 weight & 0 & $\frac{3m}{4}$ & $m$ & 0 & 0 & $\frac{m}{2}$ & $\frac{3m}{4}$ & 0
\end{tabular}
\end{center}
This satisfies the promise, because all non-zero strings have weight at least $m/2$. 

\begin{figure}
\centering
\includegraphics[scale=.5]{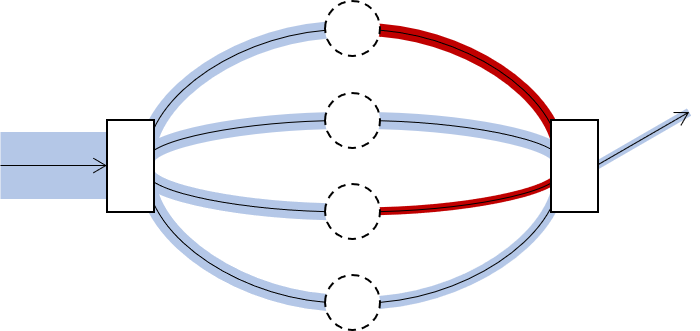}
\caption{This image depicts the probability flow through the Deutsch-Jozsa algorithm from \fig{DJ}, if we instantiate the queries with the subroutine for $h$ from \fig{h-tree}, and pause the subroutine after one step. We are assuming the different inputs to $h$ have different weights, all at least $m/2$, resulting in different amplitudes coming out of the subroutine after one step, all at least half. However, since these amplitudes are not \emph{the same}, even though two are positive and two are negative, they do not perfectly cancel, so there is non-zero amplitude on the accepting terminal.}\label{fig:DJ-composed}
\end{figure}

Consider what happens after we've run one step of the algorithm for $h$ -- a step being two queries, one for each half of the input to $h$. This is depicted in \fig{DJ-composed}. In each of the four branches of the superposition, the subroutine for $h$ has terminated with probability at least half, which we depict by some amplitude coming out of the subroutines (the remaining amplitude, not shown, is still inside the subroutine). The size of these amplitudes depends on the weight of the non-zero string in the input to $h$. In the top branch of the superposition, the input to $h$ is $x^{(1)}=0^mx^{(1,R)}$. Since $|x^{(1,R)}|=3m/4$, with probability $3/4$, we find a 1 in the first step, and conclude that $h(x^{(1)})=1$, since the one is in $x^{(1,R)}$, the right half. We thus change the amplitude to negative, shown as red -- since the subroutine has zero error, only red amplitude will ever exit the subroutine call in this branch. In the second branch, since $|x^{(2,L)}|=m$, we will find a 1 in the first step with certainty, so all amplitude has already exited the subroutine after one step, and it is positive (blue). In the third branch, since $|x^{(3,R)}|=m/2$, half the amplitude exits after one step, and in the fourth branch, since $|x^{(4,R)}|=3m/4$,  $3/4$ of the amplitude exits after one step. 

The issue is that the amount of positive and negative amplitude is not the same, and so these amplitudes do not perfectly cancel, meaning there is a nonzero amplitude on the accepting terminal, even though accepting would be incorrect for this input. If we waited for all amplitudes to exit the subroutine, then we would have the same amount coming out of each, and these would cancel, leaving 0 amplitude on the accepting terminal, but if we stop early conditioned on the subroutine being finished after one step, we will have some non-zero probabaility of outputting the wrong answer. 

Note that this cannot happen in a classical algorithm -- once you have any non-zero probability on a terminal, you're safe to accept or reject accordingly, because there's no chance that later probability will arrive and cancel it out, so that it would have been zero, had you waited.

The difference between a classical zero-error algorithm and a quantum zero-error algorithm is that in the classical picture, once we have some amount of probability on an accepting terminal, it means, with certainty, that 1 is the correct answer, and once we have some nonzero probability on a rejecting terminal, it means, with certainty, that 0 is the correct answer. This is related to the fact that a randomized algorithm can always be viewed as a tree: once you get to a terminal (leaf) by some path, there is no other path that could come and change things. This is in contrast to quantum algorithms, where even if at some point you have nonzero amplitude on an accepting terminal, it doesn't mean the answer is 1 -- it could be that if you wait, negative amplitude will come and cancel the positive amplitude so that the resulting amplitude on the accepting terminal by the end of the algorithm is 0.

\section{Transducers, Purifers, and no more log Factors}\label{sec:log}

I have already mentioned that by looking at quantum algorithms as quantum random walks, we can get some nice composition results. However, the results obtained this way are limited to composing subroutines that compute classical functions, whereas a quantum subroutine might potentially map a quantum state to another quantum state, through any \emph{unitary} linear map. For this more general type of composition, we need a more general model, called \emph{transducers}~\cite{belovs2024transducers}. Moreover, composing via transducers is how we can achieve composition of bounded-error quantum algorithms without log factors. That's what we will talk about in this section. 

Like a quantum algorithm, a transducer has an associated unitary action, and a complexity. I will not give details about what a transducer is, but it has the following properties that make it a useful abstraction for quantum algorithms:
\begin{enumerate}
\item There is a mapping that takes any quantum algorithm implementing a unitary map $U$ in complexity $T$, and compiles it into a transducer for $U$ with complexity $O(T)$.
\item There is a mapping that takes any transducer for $U$ with complexity $T$, and compiles it into a quantum circuit implementing $U$ \emph{with bounded error} in complexity $O(T)$.
\item Transducers compose nicely: You can combine a transducer for some outer algorithm that makes oracle calls, with transducers for some subroutines, to get a transducer for the composed functionality. The complexity will have a nice expression, like that in~\eq{quant-avg}. 
\end{enumerate} 

Often we want a quantum algorithm that decides a function, $f:\{0,1\}^n\rightarrow\{0,1\}$, rather than implementing some arbitrary unitary. We say a unitary $U_x$ parametrized by an input $x$ decides $f$ if it maps the quantum state that is a point function on the all-0s string, $0^m$ for some $m$ (a natural starting state) to the quantum state that is a point function on the string $0^{m-1}f(x)$. A quantum algorithm decides $f$ if it implements a unitary $U$ that decides $f$. 

Often we have quantum algorithms that don't perfectly implement the desired unitary map, $U$, but rather, implement it up to bounded error. If we try to map this to a transducer, it will also only implement $U$ up to some error. 
A $\delta$-\emph{perturbed} transducer for $U$ implements a map that is $\delta$-close to $U$ in some sense of closeness. When $\delta=0$,  we say the transducer is \emph{perfect} (with respect to some $U$ -- often a $U$ that decides a function $f$). 

When we compose transducers, the perturbations add up, similar to how errors add in algorithmic composition, so we need the perturbations to be sufficiently small so that even when they add up, we can bound them well below 1. If all of our subroutines decide functions, then we can use majority voting to decrease their error sufficiently. This incurs log factors: $D$-round majority voting has a multiplicative overhead of $D$, but reduces the error to $2^{-\Theta(D)}$, so that when we transform the subroutines into transducers, the small perturbations add up to a total perturbation well below 1, and so we can turn this into a bounded-error quantum algorithm. 

However, as stated, we can avoid the log factor overhead. It turns out that there is an analogue of majority voting for transducers that decide a function, which reduces their perturbations to $2^{-\Theta(D)}$ for any $D$, with just a $O(1)$ multiplicative overhead on the complexity. This construction is called a purifer, and we will shortly give a purifier for a simplified case, to illustrate how this process works. First, it is worth emphasizing that no matter how much we reduce perturbations in transducers, when we compile them back into algorithms, there will be a small constant error. Thus, there is no way to reduce the error of an algorithm for free by turning it into a transducer, purifying it, and then turning it back into an algorithm. This trick is just powerful enough to let us avoid the log factor overhead in composing bounded-error quantum algorithms. 

The purifier presented below is much simpler than the one in~\cite{belovs2024transducers}, but only applies to a special case. In~\cite{belovs2024purifier}, we were able to generalize this simple construction to work for any bounded-error quantum algorithm for $f$, giving a purifer with better space and query overhead than the one in \cite{belovs2024transducers}. 

\paragraph{A toy problem} To give some idea of how purification works, we will consider the special case of a quantum algorithm that outputs the quantum analogue of a biased coin that gives 0 with some probability $p_0=p_0(x)$, depending implicitly on some input $x$:
\begin{equation}
\sqrt{p_0}\e_0+\sqrt{1-p_0}\e_1.
\end{equation}
This is a 2-dimensional vector, and there are squareroots over the probabilities because quantum states are $\ell_2$-normalized. We could call such a state a biased quantum coin flip. We imagine that there is some constant $\eps\in [0,1/2)$ such that one of the following two cases holds, and we want to decide which one:
\begin{equation}\label{eq:p0}
\begin{split}
\text{Rejecting Case:}\; &  p_0\geq 1-\eps\\
\text{Accepting Case:}\; & p_0 \leq \eps.
\end{split}
\end{equation}
This setting does not fully capture the setting of bounded-error quantum algorithms, as we are assuming the algorithm's output is in a known two-dimensional space, but it is the simplest to analyze. 
But before we think about this quantum problem, let us think a little bit about the classical version of this problem.

\paragraph{A classical walk on a line} If you were given a (classical) biased coin with the promise that one of the conditions in \eq{p0} holds, and you want to know which one, your instinct would likely be to flip the coin many times, and take a majority vote. 
One way to model a majority vote is as a weighted random walk on a line. You can use the coin to implement a weighted walk on a line where the edges are labelled by the integers:

\vskip10pt

\begin{center}
\begin{tikzpicture}

\draw[<-] (-5.75,0)--(-5,0); \draw (-5,0) -- (5,0); \draw[->] (5,0) -- (5.75,0);

\filldraw (0,0) circle (.08); \node at (0,.25) {$v_0$};
\filldraw (1,0) circle (.08);
\filldraw (2,0) circle (.08);
\filldraw (3,0) circle (.08);
\filldraw (4,0) circle (.08);
\filldraw[white] (4.75,.1) rectangle (6.25,-.1); \node at (5.25,0) {$\dots$};

\node at (.5,-.25) {$0$};
\node at (1.5,-.25) {$1$};
\node at (2.5,-.25) {$2$};
\node at (3.5,-.25) {$3$};
\node at (4.5,-.25) {$4$};

\filldraw (-1,0) circle (.08);
\filldraw (-2,0) circle (.08);
\filldraw (-3,0) circle (.08);
\filldraw (-4,0) circle (.08);
\filldraw[white] (-4.75,.1) rectangle (-6.25,-.1); \node at (-5.25,0) {$\dots$};

\node at (-.5,-.25) {$-1$};
\node at (-1.5,-.25) {$-2$};
\node at (-2.5,-.25) {$-3$};
\node at (-3.5,-.25) {$-4$};
\node at (-4.5,-.25) {$-5$};
\end{tikzpicture}
\end{center}
\vskip5pt

\noindent and the weight of edge $\ell$ is 
\begin{equation}
w_{\ell} = \left(\frac{1-p_0}{p_0}\right)^\ell. 
\end{equation}
In such a graph, from vertex $v_{\ell}$, whose outgoing edges are $\ell-1$ and $\ell$, the probability of moving left, $q_L$, and right, $q_R$, respectively, are:
\begin{equation*}
q_L = \frac{\left(\frac{1-p_0}{p_0}\right)^{\ell-1}}{\left(\frac{1-p_0}{p_0}\right)^{\ell-1}+\left(\frac{1-p_0}{p_0}\right)^\ell} = p_0
\;\mbox{ and }\;
q_R = \frac{\left(\frac{1-p_0}{p_0}\right)^\ell}{\left(\frac{1-p_0}{p_0}\right)^{\ell-1}+\left(\frac{1-p_0}{p_0}\right)^\ell} = 1-p_0.
\end{equation*}
Thus, you can take steps on this graph by flipping your biased coin, and moving to the left when you see a 0, and right when you see a 1. 

If you begin in $v_0$, and then use $D$ coin flips to walk for $D$ steps, you will end up to the left of $v_0$ if you see 0s a majority of the time over your $D$ coin flips, and to the right if you see 1s the majority of the time over your $D$ coin flips. Thus, your position on the line is simply a way to remember how many more 1s than 0s you've seen, so that you can decide, after $D$ steps, what the majority was.

\paragraph{A quantum walk on a line} For the quantum analogue, we need some definitions (this will be the most technical part of the article). Let $G=(V,E)$ be a weighted undirected graph, with edge weights $\{w_e\}_{e\in E}$. The \emph{total weight} of $G$ is defined
$${\cal W}(G) = \sum_{e\in E}w_e.$$
Let $s$ and $t$ be distinct vertices in $V$. A \emph{unit $st$-flow} on $G$ is a function $\theta$ on $V\times V$ such that
\begin{enumerate}
\item $\theta(u,v)=-\theta(v,u)$ for all $u,v\in V$;
\item $\theta(u,v)=0$ whenever $\{u,v\}\not\in E$;
\item for all $u\in V\setminus \{s,t\}$, $\sum_v\theta(u,v)=0$; and
\item $\sum_v\theta(s,v) = \sum_v\theta(v,t)=1$.
\end{enumerate}
We often think of a flow as water flowing through the graph from $s$ to $t$: a unit of flow enters at $s$ -- we can imagine this happening through an additional \emph{boundary edge} at $s$, which is an edge that is incident to only one vertex -- and exits at $t$ -- perhaps also along some boundary edge -- and at any other vertex $u$, the total incoming flow (flow on edges with $\theta(v,u)>0$) equals the total outgoing flow (flow on edges with $\theta(u,v)>0$). The ``probability flows'' in \sec{q-vs-r} are examples of flows from the entrance to the (sometimes more than one) terminal(s). The effective resistance between $s$ and $t$ is defined:
$${\cal R}_{s,t}(G)=\min_{\theta}\sum_{e\in E}\frac{\theta(e)^2}{w_e}$$
where the minimization runs over unit $st$-flows, and if $e=\{u,v\}$, $\theta(e)^2=\theta(u,v)^2=\theta(v,u)^2$. The effective resistance is so named because it is the resistance across $s$ and $t$ if an appropriate potential difference is applied, in a network of electrical resistors, one for each edge of $G$, with conductances given by the edge weights. This beautiful theory is explained in~\cite{doyle1984RandomWalksAndElectriNetw}. Such electrical quantities have long-known connections to the theory of random walks, one of which is the following. Let ${\cal H}_{s,t}(G)$ be the expected number of steps needed by a random walk on $G$, starting in $s$, to reach $t$ for the first time. Then the following remarkable fact is due to~\cite{chandra1996ElectricalResAndCommute}:
\begin{equation}\label{eq:commute}
{\cal H}_{s,t}(G)+{\cal H}_{t,s}(G) = 2{\cal W}(G){\cal R}_{s,t}(G). 
\end{equation}

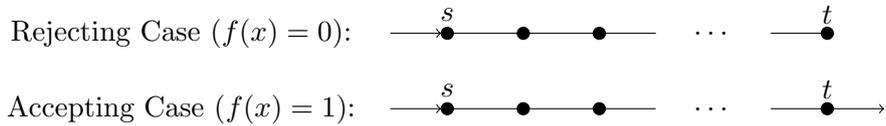
\begin{figure}[h]
\centering
\begin{tikzpicture}
\node at (-3.5,1) {Rejecting Case ($f(x)=0$):};
\node at (-3.5,0) {Accepting Case ($f(x)=1$):};

\draw[->] (-.75,1)--(-.08,1); \draw (0,1) -- (5,1);

\filldraw (0,1) circle (.08);
\filldraw (1,1) circle (.08);
\filldraw (2,1) circle (.08);
\filldraw[white] (2.75,1.1) rectangle (4.25,.9); \node at (3.5,1) {$\dots$};
\filldraw (5,1) circle (.08);

\draw[->] (-.75,0)--(-.08,0); \draw (0,0) -- (5,0); \draw[->] (5,0) -- (5.75,0);

\filldraw (0,0) circle (.08);
\filldraw (1,0) circle (.08);
\filldraw (2,0) circle (.08);
\filldraw[white] (2.75,.1) rectangle (4.25,-.1); \node at (3.5,0) {$\dots$};
\filldraw (5,0) circle (.08);

\node at (0,1.25) {$s$};	\node at (5,1.25) {$t$};
\node at (0,0.25) {$s$};	\node at (5,0.25) {$t$};
\end{tikzpicture}
\caption{The cases distinguished by a quantum walk algorithm (special case of a line).}\label{fig:qw}
\end{figure}

A quantum walk is a kind of quantum algorithm based on a weighted undirected graph. We now describe a particular type of quantum walk, which is a special case of the \emph{electric network quantum walk framework}~\cite{belovs2013ElectricWalks}. 
Consider a graph $G_x$ (depending in some way on an input $x$) with distinct vertices $s,t\in V$ that we will refer to as \emph{entrance} and \emph{terminal}. Suppose $s$ is a known vertex, perhaps labelled by $0^k$ for some $k$, to which we will always append a dangling boundary edge;
and $t$ is an unknown vertex, but we can recognize it, and we want to check if it has a certain property. If $t$ has the property -- suppose this happens precisely when $f(x)=1$ -- we will also give $t$ a dangling boundary edge, but otherwise we will not. Note that boundary edges are not really part of the graph $G$ (they are not in $E$), but we can assign them weights. We let $w_{u,\emptyset}$ be the weight of the boundary edge incident to $u$, or 0 if there is no such edge. A simple case of this setup is when $G$ is a line, as shown in \fig{qw}.

A classical random walk can distinguish these two cases in (worst-case) $\max_x{\cal H}_{s,t}(G_x)$ expected steps (which depends on the graph, but even when the graph is a line, it depends on the edge weights). 

A quantum algorithm can distinguish these two cases in the following number of steps~\cite{belovs2013ElectricWalks}:
\begin{equation}\label{eq:elec}
\sqrt{\max_{x: f(x)=0}{\cal W}(G_x)\cdot \max_{x:f(x)=1} {\cal R}_{s,t}(G_x) }.
\end{equation}
An important question is, what is meant here by step? What we mean is the quantum analogue of a classical random walk step, which is to be able to generate, for any vertex $u$, a superposition with amplitude proportional to $\sqrt{w_{u,v}}$ on the edge going out from $u$ towards $v$, which we write mathematically as
\begin{equation}\label{eq:qw-state}
\sum_{v\in V}\sqrt{w_{u,v}}\e_{u,v}+\sqrt{w_{u,\emptyset}}\e_{u,\emptyset}.
\end{equation}
Above $\e_{u,v}$ is just a vector with a $1$ in the $(u,v)$-position, and 0 everywhere else. 
A quantum operation that generates superpositions proportional to \eq{qw-state} can be used to implement a quantum operation that we call the \emph{walk operator}, and the walk operator is a transducer that decides between the accepting and rejecting cases from \fig{qw} -- that is, it decides the function $f$ -- with \emph{no perturbation}, and with complexity given by the expression in \eq{elec} (see \cite[Section 6]{belovs2024transducers}). That is, it's a \emph{perfect} transducer. 

Let us consider the complexity in \eq{elec}, and compare\footnote{The classical complexity counts the number of classical walk steps, or classical samples of an outgoing edge, whereas the quantum complexity counts the number of quantum walk steps, but for the purposes of this article, we will assume they are approximately the same difficulty to implement.} it to the classical complexity $\max_x {\cal H}_{s,t}(G_x)$. Note that \eq{elec} is upper bounded by:
$$\sqrt{\max_{x}{\cal W}(G_x)\cdot {\cal R}_{s,t}(G_x) }$$
which is equal to 
$$\sqrt{\max_{x}\frac{1}{2}(H_{s,t}(G_x)+H_{t,s}(G_x))},$$
 by \eq{commute}. This suggests that we get at best a squareroot speedup over the classical complexity. However, we will see that we can do better by maximizing separately over the two product terms.

\paragraph{Simple purification} So returning to our toy example, suppose you have a quantum subroutine that outputs a superposition $\sqrt{p_0}\e_0+\sqrt{1-p_0}\e_1=\sqrt{p_0(x)}\e_0+\sqrt{1-p_0(x)}\e_1$, the quantum version of a biased coin flip, satisfying one of the conditions in \eq{p0}. 
Let $f$ be the function that takes value $f(x)=0$ precisely when $p_0(x)\geq 1-\eps$ (the rejecting case), and $f(x)=1$ precisely when $p_0(x)\leq \eps$ (the accepting case).  

Let $G$ be the line graph of length $D$, with vertices labelled $s=v_1,\dots,v_D=t$, edges labelled $1,\dots,D-1$, entrance boundary edge (into $s$) labelled 0; terminal boundary edge (out of $t$) labelled $D$; and edge weights to be specified momentarily. 
With one call to the subroutine, we can, for any $\ell$, generate the superposition 
\begin{equation}\label{eq:qw-line}
\sqrt{p_0}\e_{\ell}+\sqrt{1-p_0}\e_{\ell+1},
\end{equation}
the quantum analog of sampling a bit, and adding $\ell$ to the sampled bit. Since the state in \eq{qw-line} is proportional to:
$$\left(\frac{1-p_0}{p_0}\right)^{\ell/2}\e_{\ell}+\left(\frac{1-p_0}{p_0}\right)^{(\ell+1)/2}\e_{\ell+1},$$
this allows us to implement a quantum walk on $G$ with edge weights set to 
$$w_\ell = \left(\frac{1-p_0}{p_0}\right)^{\ell}.$$

Now we want to distinguish between two cases:

\vskip10pt
\begin{center}
\begin{tikzpicture}
\node at (-4.3,1) {Rejecting Case, $p_0\geq 1-\eps$, $f(x)=0$:};
\node at (-4,0) {Accepting Case, $p_0\leq \eps$, $f(x)=1$:};

\draw[->] (-.75,1)--(-.08,1); \draw (0,1) -- (5,1); \draw[gray, ->] (5,1) -- (5.75,1);

\filldraw (0,1) circle (.08);
\filldraw (1,1) circle (.08);
\filldraw (2,1) circle (.08);
\filldraw[white] (2.75,1.1) rectangle (4.25,.9); \node at (3.5,1) {$\dots$};
\filldraw (5,1) circle (.08);

\draw[->] (-.75,0)--(-.08,0); \draw (0,0) -- (5,0); \draw[->] (5,0) -- (5.75,0);

\filldraw (0,0) circle (.08);
\filldraw (1,0) circle (.08);
\filldraw (2,0) circle (.08);
\filldraw[white] (2.75,.1) rectangle (4.25,-.1); \node at (3.5,0) {$\dots$};
\filldraw (5,0) circle (.08);

\node at (0,1.25) {$s$};	\node at (5,1.25) {$t$};
\node at (0,0.25) {$s$};	\node at (5,0.25) {$t$};

\node at (-.5,-.25) {\small 0};
\node at (.5,-.25) {\small $1$};
\node at (1.5,-.25) {\small $2$};
\node at (2.5,-.25) {\small $3$};
\node at (4.5,-.25) {\small $D-1$};
\node at (5.5,-.25) {\small $D$};

\end{tikzpicture}
\end{center}
\vskip5pt

In both cases, there is a boundary edge coming out of $t=v_D$, but in the rejecting case, the weight of this edge is 
$$w_D=\left(\frac{1-p_0}{p_0}\right)^D \leq \left(\frac{\eps}{1-\eps}\right)^D = 2^{-\Theta(D)},$$
so it's almost like the edge is not there. It can be formalized that while a quantum walk of the form in \fig{qw} is a perfect transducer for the associated decision problem (rejecting case vs. accepting case), the walk we are now describing is a \emph{perturbed} transducer for the associated decision problem, with perturbation $2^{-\Theta(D)}$ that can be made arbitrarily small by increasing~$D$. 

Let us now see that this transducer has $O(1)$ complexity, independent of $D$. I haven't defined the complexity of a transducer, but in the case of a quantum walk, it's just the expression in \eq{elec}. So we need to upper bound the total weight of the graph in the rejecting case, and the effective resistance in the accepting case. 

\paragraph{Rejecting case:} Suppose $p_0\geq 1-\eps$. We have:
\begin{equation}\label{eq:case0}
{\cal W}(G_x) = \sum_{e\in E(G)}w_e = \sum_{\ell=1}^{D-1}\left(\frac{1-p_0}{p_0}\right)^\ell \leq  \sum_{\ell=1}^{D-1}\left(\frac{\eps}{1-\eps}\right)^\ell \leq \frac{1}{1-\frac{\eps}{1-\eps}} = O(1).
\end{equation}

\paragraph{Accepting case:} Suppose $p_0\leq \eps$. Since the graph is a path, there is only one possibility for a flow: $\theta(v_{\ell},v_{\ell+1})=1$ on all edges $\ell$ between $s=v_1$ and $t=v_D$. Thus:
\begin{equation}\label{eq:case1}
{\cal R}_{s,t}(G_x) \leq \sum_{e\in E(G)}\frac{\theta(e)^2}{w_e} = \sum_{\ell=1}^{D-1}\frac{1}{w_\ell} = \sum_{\ell=1}^{D-1}\left(\frac{p_0}{1-p_0}\right)^\ell \leq  \sum_{\ell=1}^{D-1}\left(\frac{\eps}{1-\eps}\right)^\ell = O(1).
\end{equation}

\paragraph{Complexity} Together \eq{case0} and \eq{case1} imply that 
$$\sqrt{\max_{x: p_0(x)\geq 1-\eps}{\cal W}(G_x)\cdot \max_{x:p_0(x)\leq \eps} {\cal R}_{s,t}(G_x) } = O(1).$$
As I claimed that the left-hand side is the complexity of the perturbed transducer, we see that it is constant, even when $D$ is arbitrarily large. 

Note that in the  rejecting case, since $p_0 > 1-p_0$, the effective resistance blows up to $2^{\Theta(D)}$, but fortunately, we only care about the resistance in the  accepting case. Similarly, in the  accepting case, since $1-p_0>p_0$, the total weight blows up to $2^{\Theta(D)}$, but we only care about the total weight in the  rejecting case. 

\section{Final Thoughts}

While efficient methods of composition are certainly useful in practice, I believe there is also something we can learn about the power of quantum algorithms relative to classical algorithms by comparing their different composition capabilities. The picture of what that is is not yet fully clear. However, I believe we also have something to learn about the right way of abstracting programs for future quantum computers.

Quantum algorithms are not classical algorithms, and we have probably been trying to force them unnaturally into the shape of classical algorithms for too long. At the same time, the temptation to do so is understandable. We understand classical algorithms pretty well, relative to quantum algorithms, where things seem very mysterious, dramatic speedups are few and far between, and we don't really understand for the most part why fast quantum algorithms are fast (at least, not well enough to produce more fast quantum algorithms). So a model that keeps some classical intuition is desirable. My two cents is that we've been trying to hold onto the wrong classical intuition -- that of circuits -- and should borrow more heavily from the intuition of randomized algorithms, to understand both the similarities and differences between quantum and classical algorithms. 

Transducers give us a way to reason about quantum algorithms that seems more natural for them as \emph{quantum algorithms}, rather than some quantum version of a \emph{classical algorithm}. If we think about the special case of quantum walks, which are a sort of easy-to-visualize transducer, we have seen how a quantum walk lets us model a quantum algorithm, even one with error (which is most of them), as \emph{perfect} or at least \emph{arbitrarily close to perfect} objects that easily compose. I would like to humbly put forth that perhaps transducers are the more quantum model for reasoning about quantum algorithms that we have been missing.

\paragraph{Acknowledgements} I would like to thank Aleksandrs Belovs, Ronald de Wolf and Ben Lee Volk for helpful comments and suggestions on a draft of this article. This work is supported by ERC (ASC-Q) 101040624, NWO OCENW.Klein.061, NWA-ORC 1389.20.241, and the CIFAR QIS program. 

\small






\bibliographystyle{alpha}
\bibliography{refs}

\newcommand{\etalchar}[1]{$^{#1}$}
\newcommand{\lName}{1}\newcommand{\arxiv}[1]{arXiv:
  \href{https://arxiv.org/abs/#1}{\ttfamily{#1}}\removefirstdot}\newcommand{\arXiv}[1]{arXiv:
  \href{https://arxiv.org/abs/#1}{\ttfamily{#1}}\removefirstdot}\def\removefirstdot#1{\if.#1{}\else#1\fi}\providecommand{\multiletter}[1]{#1}\renewcommand{\multiletter}[1]{#1}\DeclareRobustCommand{\dutchPrefix}[2]{#2}\providecommand{\dutchPrefix}[2]{#2}\renewcommand{\dutchPrefix}[2]{#2}\newcommand{\skp}[3]{#2}\newcommand{\focs
  }[1]{\if\lName1\skp{ }{Proceedings of the #1 {IEEE} Symposium on Foundations
  of Computer Science ({FOCS})}{ }\else{FOCS}\fi}\newcommand{\stoc
  }[1]{\if\lName1\skp{ }{Proceedings of the #1 {ACM} Symposium on the Theory of
  Computing ({STOC})}{ }\else{STOC}\fi}\newcommand{\soda }[1]{\if\lName1\skp{
  }{Proceedings of the #1 {ACM-SIAM} Symposium on Discrete Algorithms
  ({SODA})}{ }\else{SODA}\fi}\newcommand{\stacs }[1]{\if\lName1\skp{
  }{Proceedings of the #1 Symposium on Theoretical Aspects of Computer Science
  ({STACS})}{ }\else{STACS}\fi}\newcommand{\itcs }[1]{\if\lName1\skp{
  }{Proceedings of the #1 Innovations in Theoretical Computer Science
  Conference (ITCS)}{ }\else{ITCS}\fi}\newcommand{\fsttcs }[1]{\if\lName1\skp{
  }{Proceedings of the #1 International Conference on Foundations of Software
  Technology and Theoretical Computer Science (FSTTCS)}{
  }\else{FSTTCS}\fi}\newcommand{\mfcs }[1]{\if\lName1\skp{ }{Proceedings of the
  #1 International Symposium on Mathematical Foundations of Computer Science
  ({MFCS})}{ }\else{MFCS}\fi}\newcommand{\ccc }[1]{\if\lName1\skp{
  }{Proceedings of the #1 {IEEE} Conference on Computational Complexity
  ({CCC})}{ }\else{CCC}\fi}\newcommand{\isit }[1]{\if\lName1\skp{ }{Proceedings
  of the #1 {IEEE} International Symposium on Information Theory ({ISIT})}{
  }\else{ISIT}\fi}\newcommand{\colt }[1]{\if\lName1\skp{ }{Proceedings of the
  #1 Conference On Learning Theory (COLT)}{ }\else{COLT}\fi}\newcommand{\nips
  }[1]{\if\lName1\skp{ }{Advances in Neural Information Processing Systems #1
  ({NIPS})}{ }\else{NIPS}\fi}\newcommand{\aistats }[1]{\if\lName1\skp{
  }{Proceedings of the #1 International Conference on Artificial Intelligence
  and Statistics ({AISTATS})}{ }\else{AISTATS}\fi}\newcommand{\icml
  }[1]{\if\lName1\skp{ }{Proceedings of the #1 International Conference on
  Machine Learning (ICML)}{ }\else{ICML}\fi}\newcommand{\icalp
  }[1]{\if\lName1\skp{ }{Proceedings of the #1 International Colloquium on
  Automata, Languages, and Programming (ICALP)}{
  }\else{ICALP}\fi}\newcommand{\esa }[1]{\if\lName1\skp{ }{Proceedings of the
  #1 Annual European Symposium on Algorithms (ESA)}{
  }\else{ESA}\fi}\newcommand{\tqc }[1]{\if\lName1\skp{ }{Proceedings of the #1
  Conference on the Theory of Quantum Computation, Communication, and
  Cryptography (TQC)}{}\else{TQC}\fi}\newcommand{\jacm }{\if\lName1\skp{
  }{Journal of the ACM}{ }\else{J. ACM}\fi}\newcommand{\acmta }{\if\lName1\skp{
  }{ACM Transactions on Algorithms}{ }\else{{ACM} Tr.
  Alg}\fi}\newcommand{\acmtct }{\if\lName1\skp{ }{ACM Transactions on
  Computation Theory}{ }\else{ACM Tr. Comp. Th.}\fi}\newcommand{\jams
  }{\if\lName1\skp{ }{Journal of the AMS}{ }\else{J. AMS}\fi}\newcommand{\pams
  }{\if\lName1\skp{ }{Proceedings of the AMS}{ }\else{Proc.
  AMS}\fi}\newcommand{\linalgappl }{\if\lName1\skp{ }{Linear Algebra and its
  Applications}{ }\else{Lin. Alg. \& App.}\fi}\newcommand{\jalgo
  }{\if\lName1\skp{ }{Journal of Algorithms}{ }\else{J.
  Alg.}\fi}\newcommand{\jcss }{\if\lName1\skp{ }{Journal of Computer and System
  Sciences}{ }\else{J. Comp. Sys. Sci.}\fi}\newcommand{\cc }{\if\lName1\skp{
  }{Computational Complexity}{ }\else{Comp. Comp.}\fi}\newcommand{\algor
  }{\if\lName1\skp{ }{Algorithmica}{ }\else{Alg.}\fi}\newcommand{\comb
  }{\if\lName1\skp{ }{Combinatorica}{ }\else{Comb.}\fi}\newcommand{\cacm
  }{\if\lName1\skp{ }{Communications of the ACM}{ }\else{Comm.
  ACM}\fi}\newcommand{\sigart }{\if\lName1\skp{ }{SIGART Bulletin}{
  }\else{SIGART Bull.}\fi}\newcommand{\sigactn }{\if\lName1\skp{ }{SIGACT
  News}{ }\else{SIGACT News}\fi}\newcommand{\eatcsbul }{\if\lName1\skp{
  }{Bulletin of the {EATCS}}{ }\else{Bull. {EATCS}}\fi}\newcommand{\siamrev
  }{\if\lName1\skp{ }{SIAM Review}{ }\else{SIAM Rev.}\fi}\newcommand{\siamjc
  }{\if\lName1\skp{ }{SIAM Journal on Computing}{ }\else{SIAM J.
  Comp.}\fi}\newcommand{\siamjo }{\if\lName1\skp{ }{SIAM Journal on
  Optimization}{ }\else{SIAM J. Opt.}\fi}\newcommand{\siamjdm }{\if\lName1\skp{
  }{SIAM Journal on Discrete Mathematics}{ }\else{SIAM J. Disc.
  Math.}\fi}\newcommand{\siamjnum }{\if\lName1\skp{ }{SIAM Journal on Numerical
  Analysis}{ }\else{SIAM J. Num. Anal.}\fi}\newcommand{\siamjmathanal
  }{\if\lName1\skp{ }{SIAM Journal on Mathematical Analysis}{ }\else{SIAM J.
  Math. Anal.}\fi}\newcommand{\discmath }{\if\lName1\skp{ }{Discrete
  Mathematics}{ }\else{Disc. Math.}\fi}\newcommand{\das }{\if\lName1\skp{
  }{Discrete Applied Mathematics}{ }\else{Disc. App.
  Math.}\fi}\newcommand{\amatstat }{\if\lName1\skp{ }{Annals of Mathematical
  Statistics}{ }\else{Ann. Math. Stat.}\fi}\newcommand{\rms }{\if\lName1\skp{
  }{Russian Mathematical Surveys}{ }\else{Russ. Math.
  Surv.}\fi}\newcommand{\invmath }{\if\lName1\skp{ }{Inventiones Mathematicae}{
  }\else{Inv. Math.}\fi}\newcommand{\jnumber }{\if\lName1\skp{ }{Journal of
  Number Theory}{ }\else{J. Num. Th.}\fi}\newcommand{\toc }{\if\lName1\skp{
  }{Theory of Computing}{ }\else{Th. Comp.}\fi}\newcommand{\cjtcs
  }{\if\lName1\skp{ }{Chicago Journal of Theoretical Computer
  Science}{}\else{Chic. J. Th. Comp. Sci.}\fi}\newcommand{\tocsys
  }{\if\lName1\skp{ }{Theory of Computing Systems}{}\else{Theory Comput.
  Syst.}\fi}\newcommand{\quantum }{\if\lName1\skp{ }{{Quantum}}{
  }\else{Quant.}\fi}\newcommand{\cmp }{\if\lName1\skp{ }{Communications in
  Mathematical Physics}{ }\else{Comm. Math. Phys.}\fi}\newcommand{\jmp
  }{\if\lName1\skp{ }{Journal of Mathematical Physics}{ }\else{J. Math.
  Phys.}\fi}\newcommand{\rspa }{\if\lName1\skp{ }{Proceedings of the Royal
  Society A}{ }\else{Proc. Roy. Soc. A}\fi}\newcommand{\qic }{\if\lName1\skp{
  }{Quantum Information and Computation}{ }\else{Quant. Inf. \&
  Comp.}\fi}\newcommand{\physrev }{\if\lName1\skp{ }{Physical Review}{
  }\else{Phys. Rev.}\fi}\newcommand{\pra }{\if\lName1\skp{ }{Physical Review
  A}{ }\else{Phys. Rev. A}\fi}\newcommand{\prb }{\if\lName1\skp{ }{Physical
  Review B}{ }\else{Phys. Rev. B}\fi}\newcommand{\pre }{\if\lName1\skp{
  }{Physical Review E}{ }\else{Phys. Rev. E}\fi}\newcommand{\prr
  }{\if\lName1\skp{ }{Physical Review Research}{ }\else{Phys. Rev.
  Research}\fi}\newcommand{\prx }{\if\lName1\skp{ }{Physical Review X}{
  }\else{Phys. Rev. X}\fi}\newcommand{\prl }{\if\lName1\skp{ }{Physical Review
  Letters}{ }\else{Phys. Rev. Lett.}\fi}\newcommand{\njp }{\if\lName1\skp{
  }{New Journal of Physics}{ }\else{New J. Phys.}\fi}\newcommand{\prapp
  }{\if\lName1\skp{ }{Physical Review Applied}{ }\else{Phys. Rev.
  Appl.}\fi}\newcommand{\physrep }{\if\lName1\skp{ }{Physics Reports}{
  }\else{Phys. Rep.}\fi}\newcommand{\rmp }{\if\lName1\skp{ }{Reviews of Modern
  Physics}{ }\else{Rev. Mod. Phys. }\fi}\newcommand{\phystoday
  }{\if\lName1\skp{ }{Physics Today}{ }\else{Phys.
  Today}\fi}\newcommand{\physics }{\if\lName1\skp{ }{Physics}{
  }\else{Phys.}\fi}\newcommand{\nature }{\if\lName1\skp{ }{Nature}{
  }\else{Nat.}\fi}\newcommand{\natcomm }{\if\lName1\skp{ }{Nature
  Communications}{ }\else{Nat. Comm.}\fi}\newcommand{\natphys }{\if\lName1\skp{
  }{Nature Physics}{ }\else{Nat. Phys.}\fi}\newcommand{\npjqi }{\if\lName1\skp{
  }{npj Quantum Information}{ }\else{npj Quant. Inf.}\fi}\newcommand{\scirep
  }{\if\lName1\skp{ }{Scientific Reports}{ }\else{Sci.
  Rep.}\fi}\newcommand{\science }{\if\lName1\skp{ }{Science}{
  }\else{Sci.}\fi}\newcommand{\jpa }{\if\lName1\skp{ }{Journal of Physics A:
  Mathematical and Theoretical}{ }\else{J. Phys. A}\fi}\newcommand{\ijtp
  }{\if\lName1\skp{ }{International Journal of Theoretical Physics}{
  }\else{Int. J. Th. Phys.}\fi}\newcommand{\jmo }{\if\lName1\skp{ }{Journal of
  Modern Optics}{ }\else{J. Mod. Opt.}\fi}\newcommand{\jstatph
  }{\if\lName1\skp{ }{Journal of Statistical Physics}{ }\else{J. Stat.
  Phys.}\fi}\newcommand{\pnas }{\if\lName1\skp{ }{Proceedings of the National
  Academy of Sciences}{ }\else{PNAS}\fi}\newcommand{\lncs }{\if\lName1\skp{
  }{Lecture Notes in Computer Science}{ }\else{L. Notes Comp.
  Sci.}\fi}\newcommand{\lnai }{\if\lName1\skp{ }{Lecture Notes in Artificial
  Intelligence}{ }\else{L. Notes Art. Int.}\fi}\newcommand{\lnm
  }{\if\lName1\skp{ }{Lecture Notes in Mathematics}{ }\else{L. Notes
  Math.}\fi}\newcommand{\tams }{\if\lName1\skp{ }{Transactions of the American
  Mathematical Society}{ }\else{Trans. AMS}\fi}\newcommand{\ieeetit
  }{\if\lName1\skp{ }{{IEEE} Transactions on Information Theory}{ }\else{{IEEE}
  Trans. Inf. Th.}\fi}\newcommand{\iscs }{\if\lName1\skp{ }{International
  Series in Computer Science}{ }\else{Int. Ser. Comp.
  Sci.}\fi}\newcommand{\tocl }{\if\lName1\skp{ }{Theory of Computing Library}{
  }\else{Th. Comp. Lib.}\fi}
\begin{thebibliography}{B{\dutchPrefix{Wolf}{d}}W03}

\bibitem[AB09]{arora2012ComputationalComplexity}
Sanjeev Arora and Boaz Barak.
\newblock {\em Computational Complexity: A Modern Approach}.
\newblock Cambridge University Press, 2009.

\bibitem[BB19]{blais2019separation}
Eric Blais and Joshua Brody.
\newblock Optimal separation and strong direct sum for randomized query
  complexity.
\newblock In {\em \ccc{34th}}, pages 1--17, 2019.

\bibitem[Bel13]{belovs2013ElectricWalks}
Aleksandrs Belovs.
\newblock Quantum walks and electric networks.
\newblock \arxiv{1302.3143}, 2013.

\bibitem[BJ24]{belovs2024purifier}
Aleksandrs Belovs and Stacey Jeffery.
\newblock Quantum error reduction without log factors, 2024.
\newblock \arxiv{2412.00000}.

\bibitem[BJY24]{belovs2024transducers}
Aleksandrs Belovs, Stacey Jeffery, and Duyal Yolcu.
\newblock Taming quantum time complexity.
\newblock {\em Quantum}, 8(1444), 2024.
\newblock \arxiv{2311.15873}.

\bibitem[B{\dutchPrefix{Wolf}{d}}W03]{buhrman2003zeroerror}
Harry Buhrman and Ronald {\dutchPrefix{Wolf}{d}}e~Wolf.
\newblock Quantum zero-error algorithms cannot be composed.
\newblock {\em Information Processing Letters}, 87(2):79--84, 2003.
\newblock \arxiv{quant-ph/0211029}.

\bibitem[BY23]{belovs2023LasVegas}
Aleksandrs Belovs and Duyal Yolcu.
\newblock One-way ticket to {Las Vegas} and the quantum adversary.
\newblock \arxiv{2301.02003}, 2023.

\bibitem[CRR{\etalchar{+}}96]{chandra1996ElectricalResAndCommute}
Ashok~K. Chandra, Prabhakar Raghavan, Walter~L. Ruzzo, Roman Smolensky, and
  Prasoon Tiwari.
\newblock The electrical resistance of a graph captures its commute and cover
  times.
\newblock {\em Computational Complexity}, 6(4):312--340, 1996.

\bibitem[DJ92]{deutsch1992algorithm}
David Deutsch and Richard Jozsa.
\newblock Rapid solutions of problems by quantum computation.
\newblock {\em Proceedings of the Royal Society of London A}, 1907:553--558,
  1992.

\bibitem[DS84]{doyle1984RandomWalksAndElectriNetw}
Peter~G. Doyle and J.~Laurie Snell.
\newblock {\em Random walks and electric networks}.
\newblock Mathematical Association of America, 1984.
\newblock \arxiv{math/0001057}.

\bibitem[IZ89]{impagliazzo1989recyleRandomBits}
Russell Impagliazzo and David Zuckerman.
\newblock How to recycle random bits.
\newblock In {\em \focs{30th}}, pages 248--253, 1989.

\bibitem[Jef24a]{jeffery2022subroutines}
Stacey Jeffery.
\newblock Quantum subroutine composition.
\newblock {\em Theory of Computing}, 2024.
\newblock To appear. \arxiv{2209.14146}.

\bibitem[Jef24b]{jeffery2024sigact}
Stacey Jeffery.
\newblock {SIGACT} {N}ews {C}omplexity {T}heory {C}olumn 123: Composing quantum
  algorithms.
\newblock {\em {ACM SIGACT} News}, 55(4):49–69, December 2024.

\bibitem[Rei09]{reichardt2009span}
Ben~W. Reichardt.
\newblock Span programs and quantum query complexity: The general adversary
  bound is nearly tight for every boolean function.
\newblock In {\em \focs{50th}}, pages 544--551, 2009.

\end{thebibliography}
\end{document}